\documentclass[letterpaper,10pt]{article} 
\usepackage{opticameet3} %% use version 3 for proper copyright statement

\newcommand\authormark[1]{\textsuperscript{#1}}

\usepackage{amsmath,amssymb}
\usepackage{multicol}
\usepackage{subcaption}
\usepackage[colorlinks=true,bookmarks=false,citecolor=blue,urlcolor=blue]{hyperref} %pdflatex
\begin{document}

\title{Multi-Modal Fiber Sensing for Offshore Environmental and Infrastructure Monitoring}
\vspace{-0.75 ex}
% \author{Author name(s)}
% \address{Author affiliation and full address}
% \email{e-mail address}

\author{Konstantinos Alexoudis\authormark{1,2} Florian Azendorf\authormark{1}, Alvaro Doval\authormark{3}, Steinar Bjørnstad\authormark{3}, Jasper Müller\authormark{1}, Vincent Sleiffer\authormark{1}, Chigo Okonkwo\authormark{2}, Tom Bradley\authormark{2}}

\address{\authormark{1}Adtran, Frauenhoferstraße 9a, 82152 Planegg, Germany\\
\authormark{2}High-Capacity Optical Transmission Laboratory, Eindhoven University of Technology, Netherlands\\
\authormark{3}Tampnet AS, Jåttåvågveien 7, 4020 Stavanger, Norway}

\email{\authormark{*}konstantinos.alexoudis@adtran.com} %% email address is required

%% Do not add a copyright statement. Optica will add it.

\vspace{-1.25ex}
\begin{abstract}
Monitoring a 118 km subsea cable using Distributed acoustic, state-of-polarization, and Brillouin sensing captured storm-induced strain up to $\approx$0.003 (dynamic) $\mu\epsilon$ and $\approx$180 $\mu\epsilon$ (static), demonstrating consistent yet distinct modal responses to environmental loading.
\end{abstract}

\vspace{-0.25ex}
\section{Introduction}
\vspace{-1ex}
Optical fiber cables form the backbone of today's global communication networks. In addition, they are being investigated by numerous researchers worldwide as distributed environmental sensors \cite{lu_distributed_2019}. Subsea telecom cables provide long and stable baselines, ideal for monitoring offshore and human activity.

Building on this concept, distributed fiber sensing can be realized using different modalities, such as distributed acoustic sensing (DAS), state-of-polarization (SOP), and Brillouin optical time-domain reflectometry (BOTDR) which provide different sensitivities, spatial and temporal resolutions. DAS converts standard fibers into dense vibration sensor arrays to detect earthquakes, ocean waves, and vessels \cite{landro_sensing_2022}. SOP monitoring captures polarization fluctuations linked to environmental disturbances and has been shown to reveal seismic and oceanic activity in live telecom networks without additional hardware \cite{mecozzi_polarization_2021}, while BOTDR measures the distribution of static strain and temperature along a fiber by analyzing Brillouin backscattering \cite{mathew_repeaterless_2024}.

Previous studies have demonstrated storm and ocean-wave detection using DAS \cite{meule_reconstruction_2024} or SOP \cite{pelaez_storm_2025}, and hybrid DAS/BOTDR systems have been applied to terrestrial structures \cite{carver_polarization_2024}. To the best of our knowledge, this is the first coordinated deployment of DAS, SOP, and BOTDR on an operational subsea cable. Moreover, the response of Brillouin scattering to dynamic offshore events, such as storms, remains largely unexplored.

In this work, we report a continuous two-week multi-modal offshore measurement campaign combining DAS, SOP, and BOTDR on a 118 km subsea cable from Egersund, Norway, to the offshore installation Yme, which forms part of the link connecting to the UK. The study focuses on the storm "Floris" that occurred on 5 August 2025 in Norway, analyzing the response of DAS and SOP in relation to wind data and examining BOTDR strain before and after the event. This approach enables simultaneous observation of dynamic cable excitation and slower quasi-static strain variations, providing insight into the interaction between the marine environment and subsea fiber infrastructure.

\vspace{-1ex}
\section{Field Trial Setup}
\vspace{-1ex}

\begin{figure}[htb!]
    \centering
    \includegraphics[width=\textwidth]{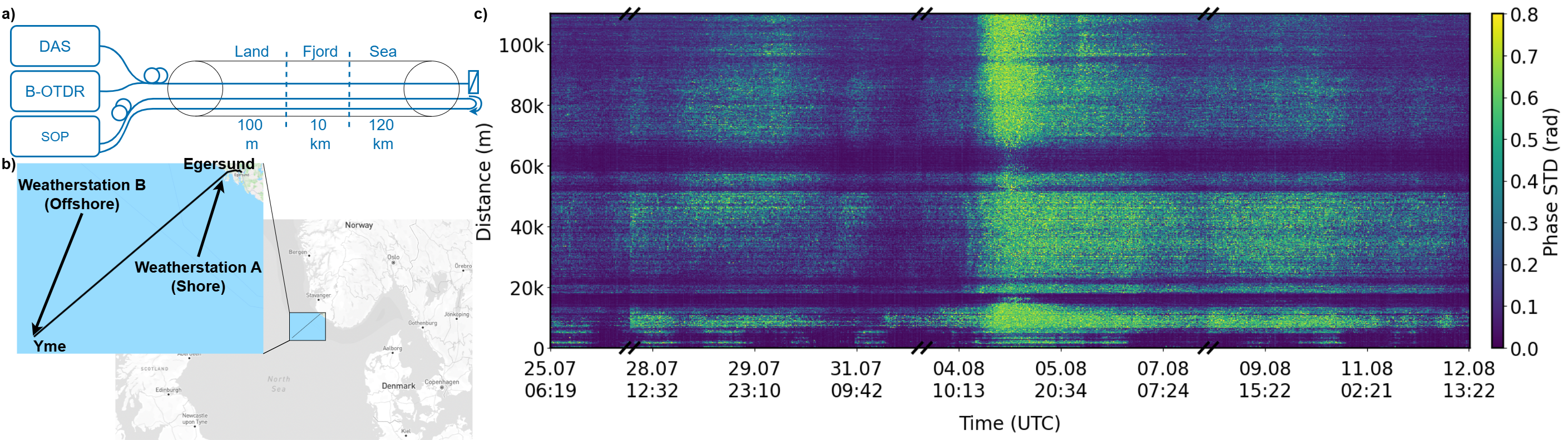}
    \vspace{-5ex}
    \caption{(a) Offshore sensing configuration combining DAS and BOTDR on one fibre and SOP monitoring on a separate fibre within the same subsea cable from Egersund, Norway, connecting to UK passing the offshore installation Yme, covering onshore, fjord, and offshore sections. (b) Map illustrating the approximate subsea cable route and the locations of the two meteorological stations referenced in this study. (c) Long-term DAS monitoring data from 25 July to 12 August 2025.}
    \vspace{-2ex}
    \label{fig:schematic}
\end{figure}

Individual fibers in a single cable between Egersund and Yme were assigned to different sensing modalities. The setup combined DAS and BOTDR on one fiber within the cable, while a second dark fiber was dedicated to SOP monitoring, allowing direct comparison between the dynamic strain rate measured by DAS, the static strain detected by BOTDR, and the polarization fluctuations observed with SOP sensing.

The DAS interrogator used a narrow-linewidth laser ($< \text{1 kHz}$) at a 1550.12 nm wavelength and launched approximately +10 dBm optical power. The interrogator samples at 250 MS/s, giving us approximately 0.4 m sample spacing \cite{alexoudis_breakdown_2025}. For the field test the DAS Interrogator measured contiuously with a gauge length of 40 m, a pulse width of 500 ns and a pulse repetition frequency (PRF) of 600 Hz.

For BOTDR operation, the same optical path generated Brillouin backscattered light, from which the Brillouin frequency shift (BFS) was extracted for strain and temperature estimation. The transmitter used the same laser source, with pulse widths between 2.5~$\mu$s and 16~$\mu$s, corresponding to nominal spatial resolutions of 0.25 km to approximately 1.6~km. The broader pulses were chosen to analyze the maximum reach of the system in deployed cables. Afterwards, to detect strain events at the shore and 20 km of cable in the sea, narrow pulses were used to increase the spatial resolution to resolve potential strain fluctuations. The BFS was measured over a frequency range of 10.6-10.67~GHz with 1~MHz steps, and each spectrum was averaged 4000 to 6000 times to improve the signal-to-noise ratio, taking 3-4 minutes. The backscattered light was mixed with a frequency-tunable local oscillator in a coherent receiver and sampled at 250 MS/s.

The SOP system used a wavelength-specific DWDM laser operating at 1550.12~nm, EDFA amplified to 22~dBm, launched into a separate fiber. The received signal was 100-GHz bandpass filtered to suppress ASE noise, and analyzed using a polarization beam splitter and two photodiodes detecting orthogonal components with a 2~Hz first-order high-pass filter at the ADC that allows to resolve 0.2~Hz at -20~dB. The vertical and horizontal components of the SOP were sampled at 44.1~kHz \cite{bjornstad_identifying_2024}.

Fig. \ref{fig:schematic}b shows the approximate cable route and meteorological stations. Fig. \ref{fig:schematic}c  shows the approximate deployed fiber route and the locations of two meteorological stations, and Fig. \ref{fig:schematic}c the analyzed period from 25 July to 12 August 2025 including storm "Floris" on 5 August 2025, which served to compare the modalities under dynamic offshore conditions.

This configuration enables direct comparison of the different modalities. SOP sensing reveals distributed activity but without localization; DAS provides spatially resolved dynamic strain information; and BOTDR adds context through quasi-static strain and temperature changes, linking transient events to long-term mechanical effects. % along the subsea cable.

\vspace{-1ex}
\section{Results}
\vspace{-1ex}

\begin{figure}[htb!]
    \centering
    \includegraphics[width=\linewidth]{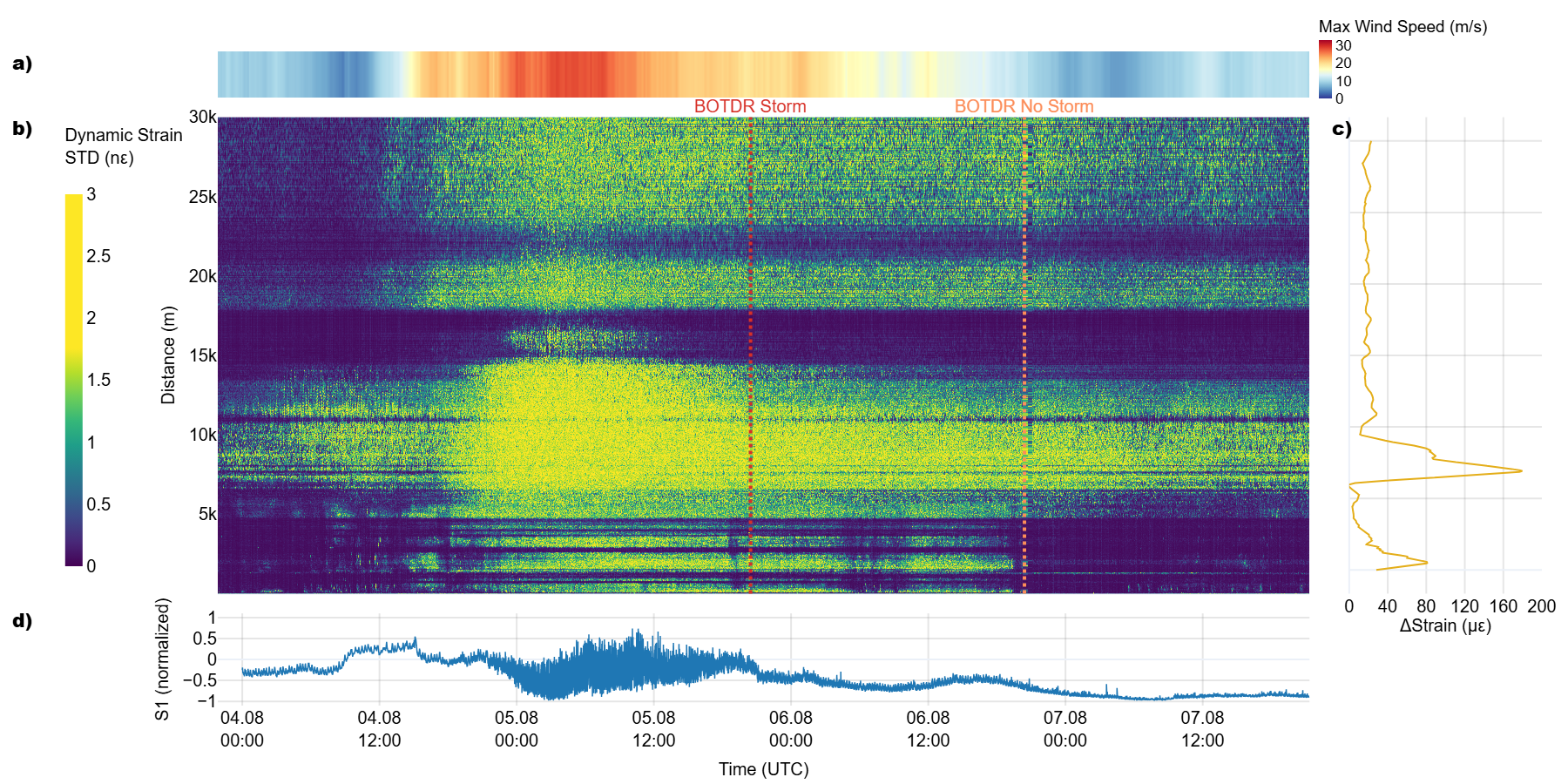}
    \vspace{-5ex}
    \caption{(a) Interpolated and averaged wind speed from two meteorological stations located near the two cable ends. (b) Downsampled standard deviation of the DAS-measured relative phase change, highlighting increased activity during the storm. (c) Difference between the BFS profiles from two BOTDR measurements taken during and after the storm, revealing local strain increases. (d) SOP variation before, during, and after the storm.}
    \vspace{-2ex}
    \label{fig:multi_plot}
\end{figure}

Fig. \ref{fig:multi_plot} summarizes the measurements acquired during the storm "Floris". Fig. \ref{fig:multi_plot}a shows the maximum wind speed obtained by interpolating and averaging data from coastal and offshore weather stations between 4 and 7 August 2025. The wind speed rose from 6 m/s on the morning of 4 August, reaching a maximum of 29 m/s on 5 August, before gradually declining and returning to normal levels at noon on 6 August. 
Fig. \ref{fig:multi_plot}b shows the DAS derived dynamic strain standard deviation in $\mu\epsilon$, calculated from the relative phase change along the fiber. Before the storm, the dynamic strain remains below 0.2 $n\epsilon$ within the first 5 km of the cable, indicating a mechanically stable near-shore section. Similarly, regions with weaker mechanical coupling, such as around 15 km, showed little variation in strain. As the storm intensified, the fjord region showed persistently high strain throughout the event, while the better protected cable sections at 15 km only showed a high dynamic strain rate during the peak of the storm. The strongest dynamic response of up to 2.9 $n\epsilon$ occurred between 5 km and 15 km, where the cable transitions from the fjord to the open sea, reflecting an enhanced mechanical coupling to environmental load.
The vertical dashed lines inside Fig. \ref{fig:multi_plot}b indicate the timing of the two BOTDR measurements, while Fig. \ref{fig:multi_plot}c presents the corresponding strain difference $\Delta\epsilon$, aligned distance-wise with the DAS data. The comparison reveals a local static-strain increase of up to $\approx82\mu\epsilon$ at 0.75 km, located in the first fjord section, and $\approx180\mu\epsilon$ near 7 km, at the fjord exit. $\Delta\epsilon$ increases from 6.0 to 6.7 km, followed by a slight decrease and a plateau between 7.6 and 8.3 km before easing to 9.5 km. These strain variations match with the increased dynamic activity observed by the DAS in these sections, suggesting that the fjord region experiences enhanced mechanical coupling, resulting in both localized and distributed strain accumulating during the storm. 
Fig. \ref{fig:multi_plot}d shows the $S_1$ stokes parameter normalized over $S_0$. The absolute Stokes parameters were obtained from the RMS of the relative AC coupled signals, temporally aligned with the DAS data. The SOP response represents an accumulated effect over the entire fiber link, showing increased polarization fluctuations during the strongest wind period. In contrast, the DAS reveals where along the fiber the dynamic strain activity was most pronounced.

\begin{figure}[htb!]
    \centering
    \includegraphics[width=\textwidth]{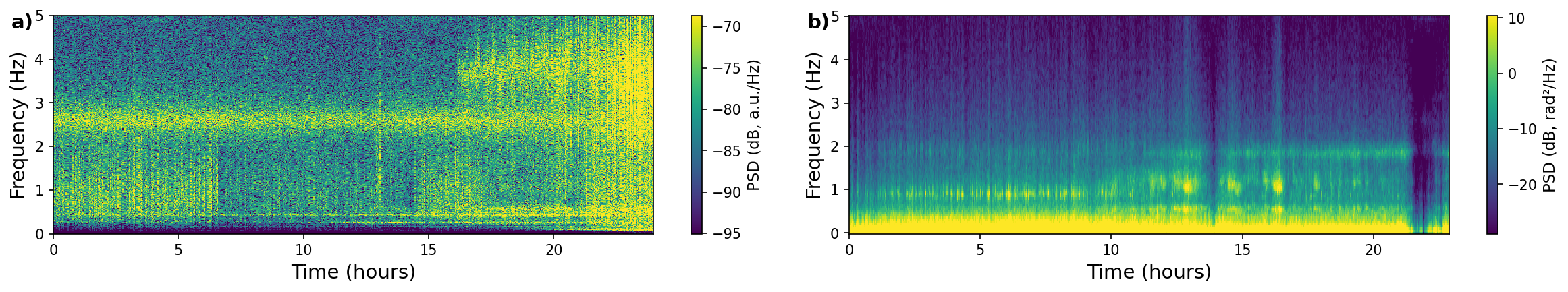}
    \vspace{-5ex}
    \caption{Spectrograms of (a) SOP and (b) DAS measurements around the storm period. Both show a narrowband component between 2 and 2.5~Hz, most prominent during the storm, illustrating the complementary sensitivity of the two sensing modalities.}
    \vspace{-2ex}
    \label{fig:spectrograms}
\end{figure}

Fig. \ref{fig:spectrograms} compares the frequency domain of the SOP and DAS data around the storm period. A narrowband component at approximately 2 to 2.5 Hz was observed in both datasets. The small frequency offset between modalities is likely related to their differing sensitivities and working mechanisms where DAS measures local dynamic strain, whereas SOP integrates birefrigence and rotation fluctuations throughout the fiber path. Analyzing the DAS data revealed multiple locations where this component originated, mainly in the first kilometers in the fjord and the first kilometers of the offshore section. Although the offshore section was also visible before the storm, the frequency was only visible in the Fjord section during the storm and persisting at lower intensity afterwards, before disappearing, suggesting that the storm amplified standing wave like oscillations in this section.

\vspace{-1.5ex}
\section{Conclusion}
\vspace{-1ex}
The combined measurements highlighted the strength of multi-modal sensing utilizing SOP to obtain a summarized view of events, BOTDR to gain structural insights and finally DAS to detect spatially resolved vibration events. During storm "Floris", SOP captured polarization fluctuations reflecting the overall mechanical excitation along the 118 km fiber. DAS localized these excitations with dynamic amplitudes of up to 2.9 $n\epsilon$ and localized specific regions that are more strongly coupled to the environment. BOTDR revealed, that only some of these dynamically active regions exhibited static strain offsets of up to $\approx$180 $\mu\epsilon$, which largely relaxed after the storm, indicating a temporary redistribution of mechanical stress rather than a lasting deformation, potentially influenced by cable slack. The 2-2.5 Hz response observed in both SOP and DAS indicates coherent responses between the different modalities, with their differing spatial and temporal resolutions offering complementary insights. %Future work will investigate the coupling dynamics between short-lived strain disturbances and quasi-static fiber tension.
\newline
\begin{footnotesize}
\textbf{Acknowledgements:} This work has received funding from the Horizon Europe Framework Programme under grant agreement No 101189703 (ICON Project). We also acknowledge the Bilateral Project "DistraSignalSense" between the TU Eindhoven, The Netherlands, and Adtran Networks SE.
\end{footnotesize}

\end{document}